\documentstyle[12pt,psfig,aasms4]{article}
\def\degr{\hbox{$^\circ$}}
\def\arcmin{\hbox{$^\prime$}}
\def\arcsec{\hbox{$^{\prime\prime}$}}

\def\fs{\hbox{$.\!\!^{\rm s}$}}

\def\gsim{\mathrel{\hbox{\rlap{\lower.55ex \hbox {$\sim$}}
                   \kern-.3em \raise.4ex \hbox{$>$}}}}
\def\lsim{\mathrel{\hbox{\rlap{\lower.55ex \hbox {$\sim$}}
                   \kern-.3em \raise.4ex \hbox{$<$}}}}
\newcommand{\gpm}[3]{$#1^{+#2}_{-#3}$}
\begin{document}  
\title{The X-ray, Optical and Infrared Counterpart to GRB\,980703}

\author{
P.M. Vreeswijk\altaffilmark{1}, 
T.J. Galama\altaffilmark{1},        
A. Owens\altaffilmark{2},
T. Oosterbroek\altaffilmark{2},
T.R. Geballe\altaffilmark{3},
J. van Paradijs\altaffilmark{1,4},
P.J. Groot\altaffilmark{1},
C. Kouveliotou\altaffilmark{5,6},
T. Koshut\altaffilmark{5,6},
N. Tanvir\altaffilmark{7},
R.A.M.J. Wijers\altaffilmark{8},
E. Pian\altaffilmark{9}, 
E. Palazzi\altaffilmark{9},
F. Frontera\altaffilmark{9,10},
N. Masetti\altaffilmark{9},
C. Robinson\altaffilmark{4,6},
M. Briggs\altaffilmark{4,6},
J.J.M. in 't Zand\altaffilmark{11},
J. Heise\altaffilmark{11},
L. Piro\altaffilmark{12},
E. Costa\altaffilmark{12},
M. Feroci\altaffilmark{12},
L.A. Antonelli\altaffilmark{12},
K. Hurley\altaffilmark{13},
J. Greiner\altaffilmark{14},
D.A. Smith\altaffilmark{15},
A.M. Levine\altaffilmark{15},
Y. Lipkin\altaffilmark{16},
E. Leibowitz\altaffilmark{16},
C. Lidman\altaffilmark{17},
A. Pizzella\altaffilmark{17},
H. B\"ohnhardt\altaffilmark{17},
V. Doublier\altaffilmark{17},
S. Chaty\altaffilmark{18,19},
I. Smail\altaffilmark{20},
A. Blain\altaffilmark{21},
J.H. Hough\altaffilmark{22},
S. Young\altaffilmark{23},
N. Suntzeff\altaffilmark{24}
}

\altaffiltext{1}{Astronomical Institute `Anton Pannekoek', University
of Amsterdam, \& Center for High Energy Astrophysics, Kruislaan 403,
1098 SJ Amsterdam, The Netherlands} 

\altaffiltext{2}{Astrophysics Division, Space Science Department of
ESA, European Space Research and Technology Centre, 2200 AG Noordwijk,
The Netherlands}

\altaffiltext{3}{Joint Astronomy Centre, 660 N. A'ohoku Place, Hilo, Hawaii 96720, USA}

\altaffiltext{4}{Physics Department, University of Alabama in
Huntsville, Huntsville AL 35899, USA}

\altaffiltext{5}{Universities Space Research Association}

\altaffiltext{6}{NASA/MSFC, Code ES-84, Huntsville AL 35812, USA}

\altaffiltext{7}{Institute of Astronomy, Madingley Road, Cambridge CB3 0HA, UK}

\altaffiltext{8}{Astronomy Program, State University of NY, Stony
Brook, NY 11794-3800, USA}

\altaffiltext{9}{Istituto Tecnologie e Studio Radiazioni
Extraterrestri (TESRE), CNR, Via P. Gobetti 101, 40 129 Bologna,
Italy}

\altaffiltext{10}{Dipartimento di Fisica Universita' di Ferrara, Via
Paradiso 12, 44100 Ferrara, Italy}

\altaffiltext{11}{Space Research Organisation Netherlands (SRON),
Sorbonnelaan 2, 3584 CA Utrecht, The Netherlands }

\altaffiltext{12}{Istituto di Astrofisica Spaziale, CNR, Via Fosso del
Cavaliere, Roma, I-00133, Italy}

\altaffiltext{13}{University of California at Berkeley, Space Sciences
Laboratory, Berkeley, CA, USA 94720-7450}

\altaffiltext{14}{Astrophysikalisches Institut Potsdam, D-14482
Potsdam, Germany}

\altaffiltext{15}{Massachusetts Institute of Technology, 77 Mass. Avenue, Cambridge, MA 02139-4307, USA} 

\altaffiltext{16}{Wise Observatory, Tel Aviv University, Ramat Aviv, Tel Aviv 69978, Israel}

\altaffiltext{17}{ESO, Casilla 19001, Santiago 19, Chile}

\altaffiltext{18}{DAPNIA/Service d'Astrophysique, CEA/Saclay, F-91191
Gif-Sur-Yvette, France}

\altaffiltext{19}{Centre d'Etude Spatiale des Rayonnements, 9, avenue
du Colonel Roche BP 4346, F-31 028 Toulouse Cedex 4, France}

\altaffiltext{20}{University of Durham, South Road, Durham, DH1 3LE, UK}

\altaffiltext{21}{Cavendish Laboratory, Madingley Road, Cambridge CB3
0HE, UK}

\altaffiltext{22}{Physics \& Astronomy, University of Hertfordshire, Hatfield, AL10 9AB, UK} 

\altaffiltext{23}{Cerro Tololo Interamerican Observatory, Casilla 603,
La Serena, Chile}

\begin{abstract}

We report on X-ray, optical and infrared follow-up observations of
GRB\,980703.  We detect a previously unknown X-ray source in the GRB
error box; assuming a power law decline we find for its decay index
$\alpha <$ --0.91 (3$\sigma$).  We invoke host galaxy extinction to
match the observed spectral slope with the slope expected from
`fireball' models. We find no evidence for a spectral break in the
infrared to X-ray spectral range on 1998 July 4.4, and determine a
lower limit of the cooling break frequency: $\nu_{\rm c}$ $>$ 1.3
$\times$ 10$^{17}$ Hz. For this epoch we obtain an extinction of A$_V
= 1.50 \pm 0.11$.  From the X-ray data we estimate the optical
extinction to be A$_V$ = \gpm{20.2}{12.3}{7.3}, inconsistent with the
former value. Our optical spectra confirm the redshift of $z$=0.966
found by Djorgovski et al. (1998). We compare the afterglow of
GRB\,980703 with that of GRB\,970508 and find that the fraction of the
energy in the magnetic field, $\epsilon_{B} < 6\times10^{-5}$, is much
lower in the case of GRB\,980703, which is a consequence of the high
frequency of the cooling break.

\end{abstract}


\setcounter{footnote}{0}
\section{Introduction}
\label{sec:intro}

Several properties of gamma-ray burst (GRB) afterglows can be well
explained by `fireball' models, in which a relativistically expanding
shock front, caused by an energetic explosion in a central compact
region, sweeps up the surrounding medium and accelerates electrons in
a strong synchrotron emitting shock (M\'eszar\'os and Rees 1994,
Wijers, Rees and M\'eszar\'os 1997; Sari, Piran and Narayan 1998;
Galama et al.  1998a). The emission shows a gradual softening with
time, corresponding to a decrease of the Lorentz factor of the
outflow. Most X-ray and optical/infrared (IR) afterglows display a
power law decay (except GRB\,980425, which is most likely associated
with the peculiar supernova SN\,1998bw; Galama et al. 1998b).

Spectral transitions in the optical/IR have been detected for the
afterglows of GRB\,970508 (Galama et al. 1998a; Wijers and Galama
1998) and GRB\,971214 (Ramaprakash et al.  1998). These have been
explained by the passage through the optical/IR waveband of the
cooling break, at $\nu_{\rm c}$ (for GRB\,970508) and the peak of the
spectrum, at $\nu_{\rm m}$ (for GRB\,971214); see Sari, Piran \&
Narayan (1998) and Wijers \& Galama (1998) for their definition. For
GRB\,970508 the observed break is in excellent agreement with such
`fireball' models, while for GRB\,971214 an exponential extinction has
been invoked to explain the discrepancy between the expected and
observed spectral index.

GRB\,980703 was detected on July 3.182 UT with the All Sky Monitor
(ASM) on the {\it Rossi X-ray Timing Explorer} (RXTE; Levine et
al. 1998), the {\it Burst And Transient Source Experiment} (BATSE,
trigger No. 6891; Kippen et al. 1998), {\it Beppo}\,SAX (Amati et
al. 1998) and {\it Ulysses} (Hurley et al. 1998). The burst as seen by
BATSE consisted of two pulses, each lasting approximately 100 sec.,
with a total duration of about 400 sec. (Kippen et al. 1998). The
first pulse had significant sub-structure, whereas the second, weaker
episode was relatively smooth. This double-peak morphology has also
been seen with the {\it Beppo}\,SAX {\it Gamma Ray Burst Monitor}
(Amati et al. 1998).  BATSE measured a peak flux of (1.9 $\pm$ 0.1)
$\times$ 10$^{-6}$ erg cm$^{-2}$ s$^{-1}$ (25 - 1000 keV) and fluence
of (4.6 $\pm$ 0.4) $\times$10$^{-5}$ erg cm$^{-2}$ ($>$ 20 keV),
consistent with the {\it Beppo}\,SAX GRBM measurement.  A time
resolved spectral analysis of the burst will be presented elsewhere
(Koshut et al. 1999).

Observations with the Narrow-Field Instruments (NFIs) of {\it
Beppo}\,SAX showed a previously unknown X-ray source (Galama et
al. 1998c) inside both the ASM error box and the InterPlanetary
Network annulus (Hurley et al. 1998). Frail et al. (1998a; see also
Zapatero Osorio et al. 1998) subsequently reported the discovery of a
radio (6 cm) and optical (R-band) counterpart to GRB\,980703.

Here we report X-ray (0.1-10 keV), optical (VRI), and infrared (JHK)
follow-up observations of GRB\,980703. In \S \ref{sec:nfi} we report
our NFI X-ray observations of the ASM error box, and \S \ref{sec:opir}
is devoted to the description and results of our spectroscopic and
photometric optical/IR monitoring campaign. We discuss the results of
these observations in \S \ref{sec:dis}.

\section{X-ray observations}
\label{sec:nfi}

We observed the ASM error box of GRB\,980703 with the {\it Beppo}\,SAX
Low- and Medium Energy Concentrator Spectrometers (LECS, 0.1-10 keV,
Parmar et al. 1997; MECS, 2-10 keV, Boella et al. 1997) on July
4.10-5.08 UT (starting 22 hrs after the burst) and on July 7.78-8.71
UT. The LECS and MECS data show a previously unknown X-ray source
1SAX\,J2359.1+0835 at R.A. = 23$^{\rm h}$59$^{\rm m}$06\fs8, Decl. =
+08\degr35\arcmin45\arcsec (equinox J2000.0), with an error radius of
50\arcsec. The field also contains the sources 1SAX J2359.9+0834 at R.A.
= 23$^{\rm h}$59$^{\rm m}$59$^{\rm s}$, Decl.=
+08\degr34\arcmin03\arcsec, and 1SAX J0000.1+0817 at R.A. = 00$^{\rm
h}$00$^{\rm m}$04$^{\rm s}$, Decl.=+08\degr17\arcmin14\arcsec. Both are
outside the ASM error box, do not show any variability and coincide
with the known ROSAT sources, 1RXS J235959.1+083355 and 1RXS
J000007.0+081653, respectively.

\begin{figure}

\caption[]{The 2-10 keV light curve of GRB\,980703. Time and flux are
on a logarithmic scale. \label{fig:xray}}
\centerline{\psfig{figure=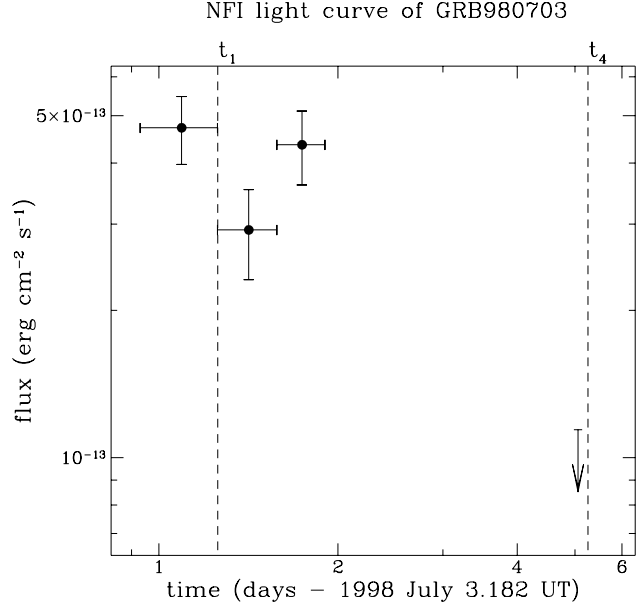,angle=0,width=8.8cm}}
\end{figure}

We extracted 1SAX J2359.1+0835 data at the best fit centroids with
radii of 8$^{'}$ (LECS) and 2$^{'}$ (MECS). To analyze the spectrum we
binned the data into channels, such that each contained at least 20
counts.  Using the separate standard background files of the
spectrometers, we simultaneously fitted the LECS and MECS data of July
4-5 UT, with a power law spectrum and a host galaxy absorption
cut-off, using a redshift of $z = 0.966$ (Djorgovski et al. 1998; see
also \S \ref{sec:opir} of this paper). In the fit we fixed the
Galactic foreground absorption at $N_{\rm H}$ = 3.4 $\times$ 10$^{20}$
cm$^{-2}$ (A$_V$ = 0.19, as inferred from the dust maps of Schlegel,
Finkbeiner \& Davis 1998\footnote{see
http://astro.berkeley.edu/davis/dust/index.html}, and the
A$_V$-N$_{\rm H}$ relation of Predehl \& Schmitt 1995). The average
spectrum can be modelled with a photon index $\Gamma = 2.51 \pm 0.32$
and a host galaxy column density $N_{\rm H}({\rm host})$ =
\gpm{3.6}{2.2}{1.3} $\times$ 10$^{22}$ cm$^{-2}$, corresponding to
A$_{V}({\rm host})$ = \gpm{20.2}{12.3}{7.3} (local to the
absorber). Modelling the spectrum without $N_{\rm H}({\rm host})$
results in a very poor fit. We did not account for a possible small
position dependent error in the relative flux normalizations between
the LECS and MECS, which is only a few percent near the center of the
image.  A change of 25\% in the assumed Galactic foreground absorption
does not affect the output values of the fit parameters.  For the
second epoch of NFI observations we kept the position fixed at the
position determined from the first epoch; we find F$_{X} <$ 1.1
$\times$ 10$^{-13}$ erg cm$^{-2}$ s$^{-1}$ (3$\sigma$). Fitting a
power law model to the light curve including the upper limit, we
obtain $\alpha <$ --0.91 for the decay index. The X-ray light curve is
shown in Fig \ref{fig:xray}.  We checked for the presence of the 6.4
keV K line at the redshifted energy of 3.26 keV, but do not detect
it. The upper limit on its flux is 8.3 $\times$ 10$^{-6}$ photons
cm$^{-2}$ s$^{-1}$ (90\% confidence level), corresponding to an
equivalent width of 532 eV in the observer's frame.

\section{Optical and infrared observations}
\label{sec:opir}

We observed the field of GRB\,980703 with the Wise Observatory 1-m
telescope (in I); the 3.5-m New Technology Telescope (NTT; in V, I and
H), the 2.2-m (in H and Ks) and Dutch 90-cm (in gunn i) telescopes of
ESO (La Silla); the CTIO 0.9-m telescope (in R) and UKIRT (in H, J and
K).

The optical images were bias-subtracted and flat-fielded in the
standard fashion. The infrared frames were reduced by first removing
bad pixels and combining about five frames around each object image to
obtain a sky image. This sky image was then subtracted after scaling
it to the object image level; the resulting image was flat-fielded.
Four reference stars were used to obtain the differential magnitude of
the optical transient (OT) in each frame. These stars were calibrated
by observing the standard stars PG2331+055 (in V and I; Landolt 1992),
FS2 and FS32 (in J, H and K; Casali \& Hawarden 1992). We used the
R-band calibration of Rhoads et al. (1998). The offsets in right
ascension and declination from the OT, and the apparent standard
magnitudes outside the Earth's atmosphere of the reference stars are
listed in Table \ref{tab:refstars}.

\begin{figure}

\caption[]{V, R, I, J, H and K light curves of GRB\,980703. The filled
symbols denote our data, while the open symbols represent data taken
from the literature (Zapatero Osorio et al. 1998; Rhoads et al. 1998;
Henden et al. 1998; Bloom et al. 1998; Pedersen et al. 1998;
Djorgovski et al. 1998; Sokolov et al. 1998 \& private communication).
For each filter a power law model plus a constant: $F_{\nu} = F_{0}
\cdot t^{\alpha} + F_{\rm gal}$ is fitted (solid lines). The fit
parameters are listed in Table \ref{tab:mag}. The times t$_1$ - t$_4$, at
which we have reconstructed the spectral flux distribution of the OT,
are indicated by the dashed lines. \label{fig:mags.ps}}
\centerline{\psfig{figure=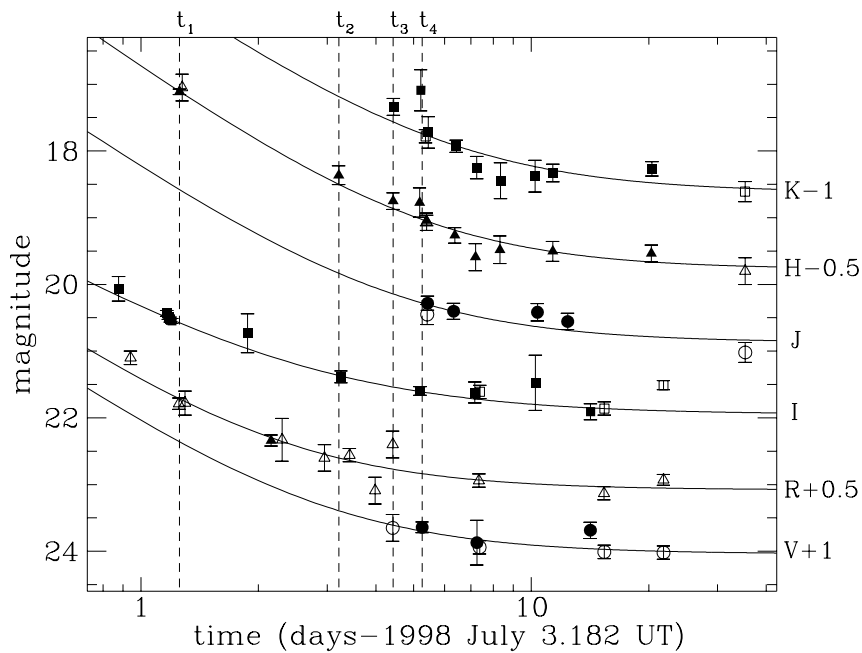,angle=-90,width=8.8cm}}
\end{figure}

The light curves of the OT are shown in Fig. \ref{fig:mags.ps} (see
Table \ref{tab:log} for a list of the magnitudes). In view of the
flattening of the light curves after $t \sim 5$ days we fitted a model
$F_{\nu} = F_{0} \cdot t^{\alpha} + F_{\rm gal}$ to our own I and H
band light curves (in these bands we have sufficient data for a free
parameter fit). Here $t$ is the time since the burst in days, and
$F_{\rm gal}$ is the flux of the underlying host galaxy.  The values
for $m_{0}= -2.5 \cdot {\rm log}\,(F_{0})+C$, the decay index $\alpha$
and $m_{\rm gal}= -2.5 \cdot {\rm log}\,(F_{\rm gal})+C$, are listed
in Table \ref{tab:mag}. The photometric calibration, which determines
$C$, has been taken from Bessell (1979) for V, R and I and Bessell \&
Brett (1988) for J, H and K.  The weighted mean value of $\alpha$ for
the I and H-bands equals $-1.61 \pm 0.12$, while an I and H-band joint
fit, with a single power law decay index, gives $\alpha = -1.63 \pm
0.12$ ($\chi$ = 15.4/16). For the V, R, J and K bands we fixed
$\alpha$ at --1.61, included also data from the literature, and fitted
$m_0$ and $m_{\rm gal}$. The fits are shown as solid lines in
Fig. \ref{fig:mags.ps}. We note that when we fit all three parameters
to the R band data (mainly data from the literature) we obtain a
temporal slope of $-1.94 \pm 0.22$, consistent with the adopted value
of $-1.61 \pm 0.12$. However, we do not include this value in the
average, since the R band magnitudes are taken from the literature and
thus not consistently measured.

For four epochs we have reconstructed the spectral flux distribution
of the OT (times t$_1$, t$_2$, t$_3$ and t$_4$ in Fig. \ref{fig:mags.ps},
corresponding to 1998 July 4.4, 6.4, 7.6 and 8.4 respectively). The
host galaxy flux, obtained from the fits to the light curves, was
subtracted. We corrected the OT fluxes for Galactic foreground
absorption (A$_V$ = 0.19). If more than one value per filter was
available around the central time of the epoch, we took their weighted
average. All values were brought to the same epoch by applying a
correction using the slope of the fitted light curve. For the first
epoch (t$_1$) we fitted the resulting spectral flux distribution with
a power law, $F_\nu \propto \nu^{\beta}$, and find $\beta = -2.71 \pm
0.12$.

We took three 1800 sec. spectra of the OT with the NTT, around July
8.38 UT. The \#3 grism that was used has a blaze wavelength of 4600
$\rm\AA$ and a dispersion of 2.3 \AA/pixel. The slit width was set at
1\arcsec. The three spectra were bias-subtracted and flat-fielded in
the usual way. The co-added spectrum was then extracted the same way
as the standard star Feige 110. We wavelength calibrated the spectrum
with a Helium/Argon lamp spectrum, with a residual error of 0.03
$\rm\AA$. The spectrum was flux calibrated with the standard star
Feige 110 (Massey et al. 1988). We estimate the accuracy of the flux
calibration to be 10\%. The wavelength and flux calibrated
spectrum shows one clear emission line at 7330.43 $\pm$ 0.14 $\rm\AA$
with a flux of 3.6 $\pm$ 0.4 $\times 10^{-16}$ erg cm$^{-2}$
s$^{-1}$. At the redshift $z=0.966$ determined by Djorgovski et
al. (1998) this is the $\lambda$\,3727 line of [O\,II]; our wavelength
measurement corresponds to $z=0.9665\pm 0.0005$.

\section{Discussion}
\label{sec:dis}

From the optical/IR light curves presented in \S 3 we have obtained an
average power law decay constant $\alpha$ = $-1.61\pm 0.12$.  This
value is consistent with the ones derived by Bloom et al. (1998)
($\alpha_{\rm R} = -1.22 \pm 0.35$, and $\alpha_{\rm I} = -1.12\pm
0.35$) and Castro-Tirado et al. (1999) ($\alpha_{\rm R} = -1.39 \pm
0.3$, and $\alpha_{\rm H} = -1.43\pm 0.11$).

\begin{figure}
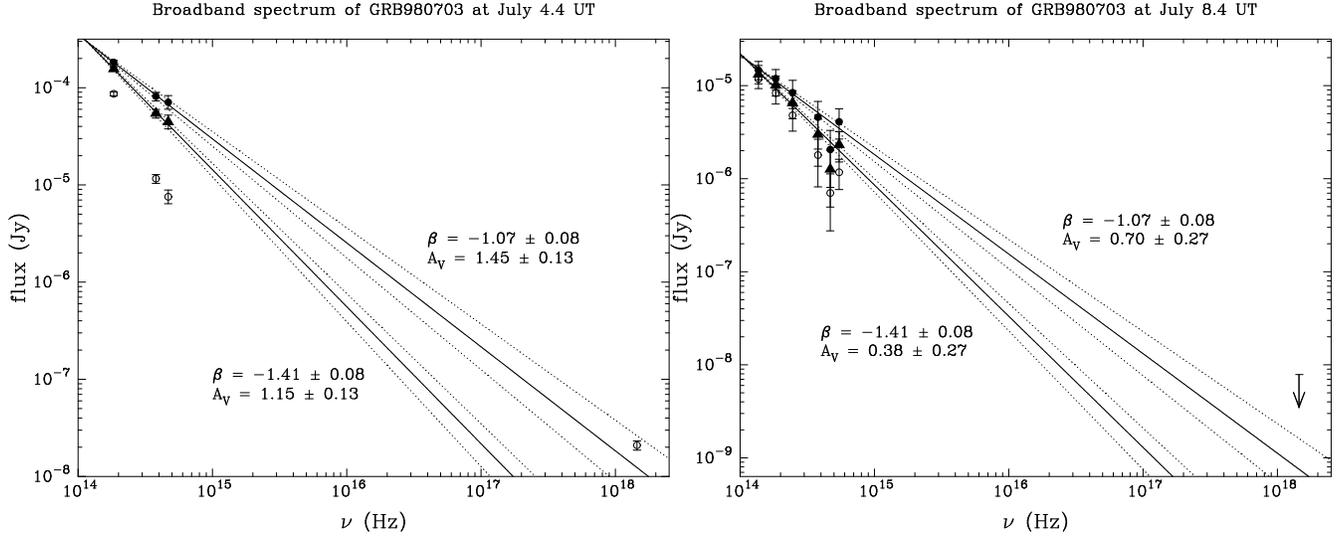

\centerline{\psfig{figure=FIG3a.ps,angle=-90,width=8.8cm}\psfig{figure=FIG3b.ps,angle=-90,width=8.8cm}}
\caption[]{Left figure: Broad-band spectrum of GRB\,980703 at July 4.4
UT (i.e., at t$_1$ in Fig. \ref{fig:mags.ps}). The open symbols are
the R, I and H OT fluxes (interpolated to July 4.4, corrected for
Galactic foreground absorption and the host galaxy flux) and the MECS
(2-10 keV) de-absorbed flux (the absorption correction is 7\%).  The
filled symbols are obtained by invoking an interstellar extinction,
A$_V$, to force the slope of the data points to take on the two
possible theoretical spectral slopes.  The two slopes $\beta$ and
their 1$\sigma$ errors are indicated by the solid and dotted lines.
Right figure: Broad-band spectrum of GRB\,980703 at July 8.4 UT (i.e.,
at t$_4$ in Fig. \ref{fig:mags.ps}). The open symbols are the V, R, I,
J, H and K OT fluxes and the MECS (2-10 keV) de-absorbed 3$\sigma$
upper limit.
\label{fig:bbspec14}}
\end{figure}

If we make the assumption that the OT emission is due to synchrotron
radiation from electrons with a power law energy distribution (with
index $p$), one expects a relation between $p$, the spectral slope
$\beta$, and the decay constant $\alpha$ (Sari, Piran and Narayan
1998). We assume that our observations are situated in the slow
cooling, low frequency regime (e.g. for GRB\,970508 this was already
the case after 500 sec.; Galama et al. 1998a). One must distinguish two
cases: (i) both the peak frequency $\nu_{\rm m}$ and the cooling
frequency $\nu_{\rm c}$ are below the optical/IR waveband.  Then $p =
(-4\alpha +2)/3 = 2.81 \pm 0.16$ and $\beta = -p/2 = -1.41 \pm 0.08$,
(ii) $\nu_{\rm m}$ has passed the optical/IR waveband, but $\nu_{\rm
c}$ has not yet. In that case $p = (-4\alpha +3)/3 = 3.15 \pm 0.16$
and $\beta = -(p-1)/2 = -1.07 \pm 0.08$.  In both cases the expected
value of $\beta$ is inconsistent with the observed $\beta = -2.71 \pm
0.12$. Following Ramaprakash et al. (1998) we assume that the
discrepancy is caused by host galaxy extinction (note that we have
already corrected the OT fluxes for Galactic foreground
absorption). To determine the host galaxy absorption we first
blueshifted the OT flux distribution to the host galaxy rest frame
(using $z$ = 0.966), and then applied an extinction correction using
the Galactic extinction curve of Cardelli, Clayton \& Mathis (1989),
to obtain the expected spectral slope $\beta$. For epoch t$_1$ (July
4.4 UT), we obtain $A_V = 1.15 \pm 0.13$ and $A_V = 1.45 \pm 0.13$ for
the cases (i) and (ii), respectively (see Fig. \ref{fig:bbspec14}).

In case (i) we find that an extrapolation of the optical flux
distribution to higher frequencies predicts an X-ray flux that is
significantly below the observed value, whereas in case (ii) the
extrapolated and observed values are in excellent agreement. The
mismatch in case (i) is in a direction that cannot be interpreted in
terms of the presence of a cooling break between the optical and X-ray
wavebands.  When we include the X-ray data point in the fit to obtain
a more accurate determination of A$_V$, we find A$_V = 1.50 \pm 0.11$,
and $\beta = -1.013 \pm 0.016$.  We conclude that the optical/IR range
is not yet in the cooling regime, and so $p$ = 3.15 $\pm$ 0.16.  Where
would the cooling frequency, $\nu_{\rm c}$, be located?  The X-ray
photon index, $\Gamma = 2.51 \pm 0.32$, corresponding to a spectral
slope of $\beta = -1.51 \pm 0.32$ suggests that perhaps the X-ray
waveband is just in the cooling regime, in which case the expected
local X-ray slope would be $\beta = -p/2 = -1.57 \pm 0.08$, while if
not in the cooling regime, it would be $\beta = -(p-1)/2 = -1.07 \pm
0.08$. However, the large error on the measured X-ray spectral slope
would also allow the cooling break to be above 2-10 keV.  We estimate
the (2$\sigma$) lower limit to the cooling frequency to be $\nu_{\rm
c}$ $>$ 1.3 $\times$ 10$^{17}$ Hz ($h \nu_{\rm c} >$ 0.5 keV).

We performed the same analysis for the other epoch (t$_4$) with X-ray
data (see Fig.  \ref{fig:bbspec14}). At this epoch, the X-ray upper
limit does not allow us to discriminate between the two
cases. However, we can still estimate a lower limit to the cooling
break from its time dependence: $\nu_{\rm c} \propto t^{-1/2}$, which
would allow the break to drop to $\nu_{\rm c}$ $>$ 6.3 $\times$
10$^{16}$ Hz only, between epoch t$_1$ and t$_4$.

On the basis of our analysis we conclude that there is no strong
evidence for a cooling break between the optical/IR and the 2-10 keV
passband before 1998 July 8.4 UT. This conclusion is at variance with
the inference of Bloom et al. (1998), who infer from their fits that
there is a cooling break at about $10^{17}$\,Hz. Upon closer
inspection, there is no real disagreement: Bloom et al. found a
slightly shallower temporal decay, and therefore a bluer spectrum of
the afterglow, which causes their extrapolated optical spectrum to
fall above the X-ray point. However, their error of 0.35 on the
temporal decay leads to an error of 0.24 on their predicted spectral
slope, and this means that a 1$\sigma$ steeper slope in their Fig. 2
would be consistent with no detected cooling break.

Assuming that the spectral slope ($\beta = -1.013 \pm 0.016$) did not
change during the time spanned by the four epochs t$_1$ - t$_4$ (as
suggested by the lack of evidence for a break in the light curve
during this timespan) we have derived the V-band extinction A$_V$ as a
function of time: A$_V$= 1.50 $\pm$ 0.11, 1.38 $\pm$ 0.35, 0.84 $\pm$
0.29 and 0.90 $\pm$ 0.25 for the epochs 1 through 4,
respectively. Fitting a straight line through these, we obtain a slope
of --0.16 $\pm$ 0.06, i.e. not consistent with zero at the 98.8\%
confidence level.  Such a decrease of the optical extinction, A$_V$,
might be caused by ionization of the surrounding medium (Perna and
Loeb 1997).

The V-band extinction A$_V$ = \gpm{20.2}{12.3}{7.3}, derived from the
host galaxy N$_H$ fit to the MECS and LECS data (July 4-5 UT) is not
in agreement with A$_V = 1.50 \pm 0.11$ as derived from the fit from
the optical spectral flux distribution. This may be due to a different
dust to gas ratio for the host galaxy of GRB\,980703, or a higher
abundance than normal of the elements that cause the X-ray absorption.

\begin{figure}
\centerline{\psfig{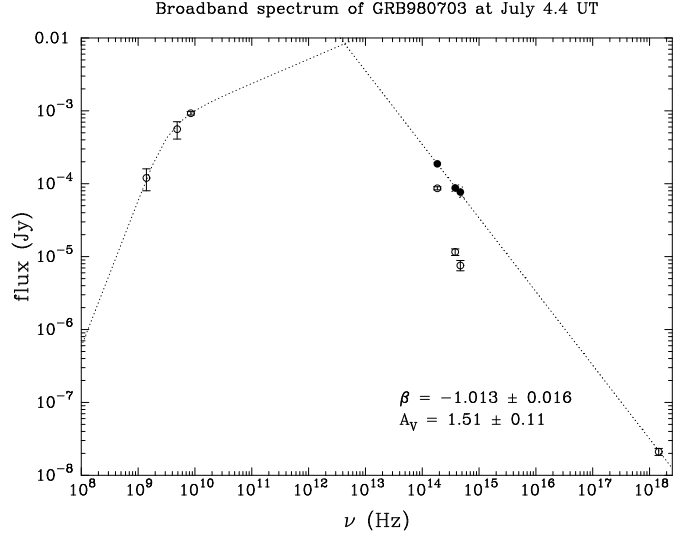}}
\caption[]{Radio to X-ray spectrum of GRB\,980703 at July 4.4 UT
(i.e., at t$_1$ in Fig. \ref{fig:mags.ps}). Shown are data from
Fig. (\ref{fig:bbspec14}) as well as 1.4, 4.86 and 8.46 GHz
observations from Frail et al. (1998b).  The fit $F_{\nu} =
F_{\nu_{\rm a}}(\nu/\nu_{\rm a})^{2}(1-\exp[-(\nu/\nu_{\rm
a})^{-5/3}])$ to the low-energy part of the spectrum with $\nu_{\rm a}
= 3.68 \pm 0.33$ GHz and $F_{\nu_{\rm a}}$ = 789 $\pm$ 42 $\mu$Jy is shown
by the dotted line. The best fit to the optical/IR and X-ray data is
also shown.
\label{fig:specfit}}
\end{figure}

With the above derived constraint on $\nu_{\rm c}$ we can partially
reconstruct the broad-band flux distribution of the afterglow of
GRB\,980703: from the radio observations of Frail et al. (1998b) at
1.4, 4.86 and 8.46 GHz, we determine the self-absorption frequency
$\nu_{\rm a}$ and its flux $F_{\nu_{\rm a}}$ from the fit $F_{\nu} =
F_{\nu_{\rm a}}(\nu/\nu_{\rm a})^{2}(1-\exp[-(\nu/\nu_{\rm
a})^{-5/3}])$ to the low-energy part of the spectrum (e.g. Granot,
Piran and Sari 1998). We have used averages of the 1.4 and 4.86 GHz
observations to obtain best estimates of the radio flux densities as
particularly those frequencies suffer from large fluctuations due to
interstellar scintillation (Frail et al. 1998b). We find $\nu_{\rm a}
= 3.68 \pm 0.33$ GHz and $F_{\nu_{\rm a}}$ = 789 $\pm$ 42
$\mu$Jy. (The fit is shown in Fig. \ref{fig:specfit}.)  The
intersection of the extrapolation from the low-frequency to the
high-frequency fit gives a rough estimate of the peak frequency,
$\nu_{\rm m} \sim 4\times10^{12}$ Hz, and of the peak flux,
$F_{\nu_{\rm m}} \sim 8$ mJy (see Fig. \ref{fig:specfit}). By assuming
such a simple broken power law spectrum the peak flux density will
likely be overestimated (realistic spectra are rounder at the peak);
it is clear from Fig. \ref{fig:specfit} that $1 < F_{\nu_{\rm m}} < 8$
mJy.

Following the analysis of Wijers and Galama (1998) we have determined
the following intrinsic fireball properties: (i) the energy of the
blast wave per unit solid angle: ${\cal E} >$ 5 $\times$ 10$^{52}$
erg/(4$\pi$ sterad), (ii) the ambient density: $n > 1.1$ nucleons
cm$^{-3}$, (iii) the percentage of the nucleon energy density in
electrons: $\epsilon_{\rm e} > 0.13$, and (iv) in the magnetic field:
$\epsilon_{B} < 6\times10^{-5}$. The very low energy in the magnetic
field, $\epsilon_B$, is a natural reflection of the high frequency of
the cooling break $\nu_{\rm c}$.

We have compared this afterglow spectrum with that of
GRB\,970508. Scaling the latter in time according to $\nu_{\rm a}
\propto t^{0}$, $\nu_{\rm m} \propto t^{-3/2}$ and $\nu_{\rm c}
\propto t^{-1/2}$, the results of GRB\,970508 (Galama et al. 1998a;
see also Granot, Piran and Sari 1998) would correspond to $\nu_{\rm a}
\sim 2.3$ GHz, $\nu_{\rm m} = 2.8\times10^{12}$ Hz, $\nu_{\rm c} =
4.8\times10^{14}$ Hz, and $F_{\nu_{\rm m}}$ = 1.3 mJy. In this
calculation we have corrected for the effect of redshift (see Wijers
and Galama 1998) such that the values represent GRB\,970508, were it
at the redshift of GRB\,980703 and observed 1.2 days after the
event. The greatest difference between the two bursts is in the
location of the cooling frequency, $\nu_{\rm c}$.

The observations on the United Kingdom Infrared Telescope, which is
operated by the Joint Astronomy Centre on behalf of the U.K. Particle
Physics and Astronomy Research Council, were carried out in Service
mode by UKIRT staff.  The BeppoSAX satellite is a joint Italian and
Dutch programme. PMV is supported by the NWO Spinoza grant.  TJG is
supported through a grant from NFRA under contract 781.76.011.  CK
acknowledges support from NASA grant NAG 5-2560. TO acknowledges an
ESA Fellowship. KH is grateful for support under JPL Contract 958056
for Ulysses, and under NASA grant NAG 5-1560 for IPN operations.

\references

Amati, L., et al. 1998, GCN circular, 146 \\
Bessell, M.S. 1979, PASP, 91, 589 \\
Bessell, M.S. \& Brett, J.M. 1988, PASP, 100, 1134 \\
Bloom, J.S., et al. 1998, \apjl, 508, L21 \\
Boella, G., et al. 1997, A\&AS, 122, 299 \\
Cardelli, J.A., Clayton, G.C. \& Mathis, J.S. 1989, \apj, 345, 245 \\
Castro-Tirado, A.J. et al. 1999, \apjl, 511, L85 \\
Casali, M.M. \& Hawarden, T.G. 1992, JCMT-UKIRT Newsletter, No. 3, 33\\
Djorgovski, S.G., et al. 1998, \apjl, 508, L17\\
Frail, D.A., et al. 1998a, GCN circular, 128 \\
Frail, D.A., et al. 1998b, GCN circular, 141 \\
Galama, T.J., et al., 1998a, \apjl, 500, L97 \\
Galama, T.J., et al. 1998b, \nat, 395, 670 \\
Galama, T.J., et al. 1998c, GCN circular, 127 \\
Granot, J., Sari, T. and Sari, R. 1998, submitted, astro-ph/9808007\\
Henden, A.A., et al. 1998, GCN circular, 131 \\
Hurley, K., et al. 1998, GCN circular, 125 \\
Kippen, M., et al. 1998, GCN circular, 143 \\
Koshut, T., et al. 1999, in preparation\\
Landolt, A.U. 1992, AJ 104, 340 \\
Levine, A., et al. 1998, \iaucirc, 6966 \\
Massey, P., et al. 1988, \apj, 328, 315\\
M\'eszar\'os, P. \& Rees, M.J. 1994, MNRAS, 269, L41 \\
Parmar, A.N., et al. 1997, A\&AS, 122, 309 \\
Pedersen, H., et al. 1998, GCN circular, 142 \\
Perna, R. \& Loeb, A. 1998, \apj, 501, 467\\
Predehl, P. and  Schmitt, J.H.M.M. 1995, A\&A, 293, 889 \\
Ramaprakash, A.N., et al. 1998, \nat, 393, 43 \\
Rhoads, J., et al. 1998, GCN circular, 144 \\
Sari, R., Piran, T., Narayan, R. 1998, \apjl, 497, L17 \\
Schlegel, D.J., Finkbeiner, D.P. and  Davis, M. 1998, \apj, 500, 525 \\
Sokolov, V., et al. 1998, GCN circular, 147 \\
Wijers, R.A.M.J. and Galama, T.J. 1998, in press, \apj \\
Wijers, R.A.M.J., Rees, M.J. and M\'eszar\'os, P. 1997, \mnras, 288,
L51 \\ 
Zapatero Osorio, M.R., et al. 1998, \iaucirc, 6967 \\

\normalsize
\begin{table}
\begin{minipage}{15cm}
\begin{tabular}{lrrrr}
\hline
filter	& star 1 		& star 2 		& star 3		& star 4 \\
\hline
$\Delta$ R.A.($\arcsec$) & --18.0	& --11.7			& --8.0			& --13.9  \\
$\Delta$ Decl.($\arcsec$) &   3.9	&  --9.7			& --21.7			& --31.8  \\
\hline
V	& 21.33 $\pm$ 0.06	& 17.02 $\pm$ 0.05 	& 22.06 $\pm$ 0.08	& 22.68	$\pm$ 0.12	\\
R	& 20.39 $\pm$ 0.04	& 16.64 $\pm$ 0.02	& 20.72 $\pm$ 0.05	& 			\\
I	& 19.55 $\pm$ 0.07 	& 16.30 $\pm$ 0.05	& 19.21 $\pm$ 0.06	& 20.01	$\pm$ 0.08	\\
J	& 18.45 $\pm$ 0.13	& 15.75 $\pm$ 0.11	& 17.52 $\pm$ 0.12	& 18.25	$\pm$ 0.13	\\
H	& 17.85 $\pm$ 0.12	& 15.44 $\pm$ 0.10	& 16.96 $\pm$ 0.11	& 17.69	$\pm$ 0.12	\\
K	& 17.71 $\pm$ 0.13	& 15.41	$\pm$ 0.12	& 16.72	$\pm$ 0.12	& 17.52	$\pm$ 0.13	\\
\end{tabular}
\end{minipage}
\caption[]{The magnitudes and offset from the OT in arc seconds of the
four comparison stars used. The error is the quadratic average of the
measurement error (Poisson noise) and a constant offset, which we
estimate to be 0.05 for the optical passbands and 0.1 for the infrared
filters. \label{tab:refstars}}
\end{table}

\scriptsize
\begin{table}
\begin{minipage}{15cm}
\begin{tabular}{rccrcl}
\hline
UT date &
magnitude &
filter &
exp. time &
seeing &
telescope/reference \\
(1998 July) &
&
&
(seconds) &
(\arcsec) & \\
\hline
  4.059 &  20.07 $\pm$  0.19 &  I & 2100 & 2.39 & Wise 1-m \\
  4.347 &  20.43 $\pm$  0.04 &  I & 900  & 1.16 & ESO NTT (EMMI) \\
  4.359 &  20.49 $\pm$  0.03 &  I & 900  & 1.10 & ESO NTT (EMMI) \\
  4.372 &  20.54 $\pm$  0.03 &  I & 900  & 1.14 & ESO NTT (EMMI) \\
  4.383 &  20.55 $\pm$  0.04 &  I & 900  & 1.02 & ESO NTT (EMMI) \\
  4.439 &  17.61 $\pm$  0.04 &  H & 810  &      & ESO NTT (SOFI) \\
  5.059 &  20.73 $\pm$  0.29 &  I & 1800 & 3.09 & Wise 1-m \\
  5.339 &  21.84 $\pm$  0.08 &  R & 3600 & 1.86 & CTIO 0.9-m \\
  6.395 &  18.86 $\pm$  0.14 &  H & 540  &       & ESO NTT (SOFI) \\
  7.609 &  19.25 $\pm$  0.12 &  H & 1200 &      & UKIRT  \\
  7.622 &  18.36 $\pm$  0.13 &  K & 600  &      & UKIRT  \\
  8.361 &  19.27 $\pm$  0.22 &  H & 2700 &      & ESO 2.2m \\
  8.375 &  21.60 $\pm$  0.06 &  I & 900  & 0.92 & ESO NTT (EMMI) \\
  8.396 &  18.14 $\pm$  0.35 &  Ks & 2700 &      &  ESO 2.2m \\
  8.438 &  22.64 $\pm$  0.08 &  V & 900  & 1.93 & ESO NTT (EMMI) \\
  8.578 &  19.54 $\pm$  0.08 &  H & 1620 &      & UKIRT  \\
  8.608 &  20.28 $\pm$  0.10 &  J & 2160 &      & UKIRT  \\
  8.633 &  18.77 $\pm$  0.24 &  K & 600  &      & UKIRT  \\
  9.509 &  20.40 $\pm$  0.12 &  J & 2160 &      & UKIRT  \\
  9.554 &  19.76 $\pm$  0.12 &  H & 2160 &      & UKIRT  \\
  9.614 &  18.94 $\pm$  0.09 &  K & 2160 &      & UKIRT \\
 10.353 &  21.62 $\pm$  0.16 &  gunn i & 4800 & 1.22 & ESO Dutch \\
 10.380 &  20.09 $\pm$  0.20 &  H & 3750 &      & ESO 2.2m \\
 10.435 &  19.24 $\pm$  0.16 &  Ks & 3900 &     &  ESO 2.2m \\
 10.440 &  22.87 $\pm$  0.34 &  V & 900  & 1.06 & ESO NTT (EMMI) \\
 11.496 &  19.98 $\pm$  0.21 &  H & 2160 &      & UKIRT \\
 11.527 &  19.47 $\pm$  0.27 &  K & 2160 &      & UKIRT \\
 13.414 &  18.76 $\pm$  0.30 &  Ks & 4950 &     &  ESO 2.2m \\
 13.438 &  21.47 $\pm$  0.41 &  gunn i & 2400 & 1.98 & ESO Dutch \\
 13.558 &  20.42 $\pm$  0.13 &  J & 2160 &      & UKIRT \\
 14.536 &  20.00 $\pm$  0.15 &  H & 2160 &      & UKIRT \\
 14.545 &  19.36 $\pm$  0.14 &  K & 2160 &      & UKIRT \\
 15.582 &  20.56 $\pm$  0.12 &  J & 3240 &      & UKIRT \\
 17.359 &  22.68 $\pm$  0.12 &  V & 900  & 1.05 & ESO NTT (SUSI2) \\
 17.371 &  21.91 $\pm$  0.12 &  I & 900  & 0.75 & ESO NTT (SUSI2) \\
 23.501 &  20.04 $\pm$  0.12 &  H & 4860 &      & UKIRT \\
 23.578 &  19.28 $\pm$  0.11 &  K & 4860 &      & UKIRT \\

\end{tabular}
\end{minipage}
\caption[]{\tiny The log of the observations with the columns:
UT Date, magnitude and error, filter, exposure time, seeing
and the telescope.

Instruments and CCDs used: NTT EMMI: red arm with TEK 2k $\times$ 2k
CCD (\#36), 0.27\arcsec /pixel; NTT SUSI2: EEV 4k $\times$ 2k CCD
(\#45 \& \#46), 0.08\arcsec /pixel; NTT SOFI: Hawaii 1k $\times$ 1k
HgCdTe array, 0.29\arcsec /pixel; ESO Dutch: CCD Camera with TEK 512
$\times$ 512 CCD (\#33), 0.47\arcsec /pixel; Wise 1-m: TEK 1k $\times$
1k CCD, 0.70\arcsec /pixel; CTIO 0.9-m: TEK 2k $\times$ 2k CCD,
0.38\arcsec /pixel; UKIRT: IRCAM3 with FPA42 256 $\times$ 256
detector, 0.29\arcsec /pixel; 2.2-m (IRAC2b): NICMOS-3 256 $\times$
256 array, 0.507\arcsec /pixel. We note that we do not list an
estimate of the seeing in case of infrared observations, since the
real seeing is overestimated due to the process of co-adding the
individual frames. \label{tab:log}}
\end{table}
\normalsize

\begin{table}
\caption[]{Fit parameters for the model $m$ = -2.5\,log$(10^{-0.4\,m_{0}}$
t$^{\alpha}$ + 10$^{-0.4\,m_{gal}})$ \label{tab:mag}}
\begin{tabular}{lllll}
\hline
filter 	& $m_{0}$		& $\alpha$ 			& m$_{gal}$ 		  & $\chi_{red}^{2}$ 	\\
\hline
V	& \gpm{21.22}{0.48}{0.33} & --1.61         		& \gpm{23.04}{0.08}{0.08} & 5.5/5		\\
R 	& \gpm{21.18}{0.09}{0.08} & --1.61        		& \gpm{22.58}{0.06}{0.05} & 14.7/10		\\
I	& \gpm{20.60}{0.04}{0.04} & \gpm{-1.36}{0.27}{0.36} 	& \gpm{21.95}{0.25}{0.16} & 4.5/8		\\
J	& \gpm{18.32}{0.33}{0.25} & --1.61        		& \gpm{20.87}{0.07}{0.11} & 5.6/4		\\
H	& \gpm{17.29}{0.06}{0.06} & \gpm{-1.67}{0.13}{0.15}     & \gpm{20.27}{0.19}{0.15} & 6.5/7		\\
K	& \gpm{16.48}{0.18}{0.15} & --1.61        		& \gpm{19.62}{0.12}{0.11} & 11.3/9		\\
\hline
\end{tabular}
\end{table}

\end{document}